\documentclass[aps,rmp,twocolumn,showpacs,a4paper]{revtex4}
\usepackage{graphicx}
 \usepackage{amsmath,amsfonts,amssymb}
\newcommand{\dd}{{\rm d}}

\newcommand{\be}{\begin{equation}}
\newcommand{\ee}{\end{equation}}
\newcommand{\ba}{\begin{eqnarray}}
\newcommand{\ea}{\end{eqnarray}}

\begin{document}
\title{Efficient and accurate three dimensional Poisson solver for surface problems}

\author{Luigi Genovese} 
\email{Luigi.Genovese@cea.fr}
\author{Thierry Deutsch} 
\affiliation{D\'epartement de recherche fondamentale sur la mati\`ere condens\'ee,\\
         SP2M/L\_Sim, CEA-Grenoble, 38054 Grenoble cedex~9, France}
\author{Stefan Goedecker}
\affiliation{Institute of Physics, University of Basel, Klingelbergstrasse 82, CH-4056 Basel, Switzerland}
\date{\today}

 \begin{abstract}
We present a method that gives highly accurate electrostatic potentials for 
systems where we have periodic boundary conditions in two spatial directions but free 
boundary conditions in the third direction. These boundary conditions are needed for all kind of surface problems. Our method has an $\mathcal O(N\log N)$ 
computational cost, where $N$ is the number of grid points, with a very small prefactor. 
This Poisson solver is primarily intended for real space methods 
where the charge density and the potential are given on a uniform grid. 
 \end{abstract}

%\pacs{ 111111111 }
%\keywords{}

 \maketitle

% main text
\section{Introduction}
Electrostatic potentials play a fundamental role in nearly any field of physics and chemistry.
Having efficient algorithms to find the electrostatic potential $V$ arising from a charge
distribution $\rho$ or, in other words, to solve the Poisson's equation 
\be \label{poissonbase}
\nabla^2 V=-4 \pi \rho\;,
\ee
is therefore essential.
The large variety of situations in which this equation can be found lead us to face this problem with different choices of the boundary conditions (BC).
The long-range behavior of the inverse Laplacian operator make this problem to be strongly dependent on the BC of the system.

The most immediate approach to the Poisson equation can be achieved for periodic BC, where a traditional reciprocal space treatment is both rapid and simple, since the Laplacian matrix is diagonal in a plane wave representation. 
If the density $\rho$ is originally given in real space, 
a first Fast Fourier Transformation (FFT) is used to transform the real space data in reciprocal space. The Poisson equation is then solved in reciprocal space and  finally the result is 
transformed back into real space by a second FFT.
Because of the FFT's, the overall computational scaling is  ($\mathcal O(N \log N)$) with respect to the number of grid points $N$. 

The situation is different if one considers the same problem for free BC.
In this case the solution of Poisson's equation can formally be obtained from a three-dimensional integral:
\be \label{greenequation}
V(\mathbf r) =\int \dd \mathbf r' G(| \mathbf r - \mathbf r' |) \rho(\mathbf r')\;,
\ee
where $G(r)=1/r$ is the Green function of the Laplacian operator in the unconstrained $\mathbb{R}^3$ space.
The long range nature of the kernel operator $G$ does not allow us to mimic free BC with a very large periodic volume.  Consequently, the description of non-periodic systems with a periodic formalism always introduces long-range interactions between super-cells that falsify the results.
Due to the simplicity of the plane wave methods, various attempts have been made to generalize the reciprocal space approach to free BC \cite{Hockney.1970, Fusti-Molnar-Pulay.2002, Martyna-Tuckerman.1999}. All of them use a FFT at some point, and have thus a $\mathcal O(N \log N)$ scaling.
These methods have some restrictions and cannot be used blindly. For example, the method by F\"usti-Molnar and Pulay is efficient only for spherical geometries, and the method by Martina and Tuckerman requires artificially large simulations boxes that are expensive numerically. Nonetheless, the usefulness of reciprocal space methods has been demonstrated for a variety of applications, and plane-wave based codes are widely used in the chemical physics community.

Another choice of the BC that is of great interest is for systems that are periodically replicated in two dimensions but with finite extent in the third, namely surface systems. The surface-specific experimental techniques developed in recent years produced important results~\cite{review}, that can benefit from theoretical prediction and analysis. The development of efficient techniques for systems with such boundary conditions thus became of great importance.
Explicit Poisson solvers have been developed in this framework \cite{Tuckerman Surfaces, Hockney Surfaces, Mortensen Surfaces}, with a reciprocal space based treatment. Essentially, these Poisson solvers are built following a suitable generalization for surfaces BC of the same methods that were developed for isolated systems. As for the free BC case, screening functions are present to subtract the artificial interaction between the super-cells in the non-periodic direction. Therefore, they exhibit the same kind of intrinsic limitations, as for example a good accuracy only in the bulk of the computational region, with the consequent need for artificially large simulation boxes which may increase the computational overhead.

Electrostatic potentials can either be calculated by solving the differential Poisson equation or 
by solving the equivalent integral equation eq.\eqref{greenequation}. 
The methods that solve the differential equation are iterative and they require
various tuning.  A good representative of these methods is the multigrid 
approach\cite{multigrid}. Several different steps such as 
\emph{smoothing}, \emph{restriction} and \emph{prolongation} are needed in this approach. 
Each of these steps has to be tuned to optimize speed and accuracy. 
Approaches based on the integral equation are in contrast straightforward and do not require 
such tuning. 

In this paper we will describe a new Poisson solver compatible with the boundary conditions 
of surfaces. Contrary to Poisson solvers based on reciprocal space treatment, 
the fundamental operations of this Poisson solver are based on a mixed reciprocal-real space 
representation of the charge density. This allows us to naturally satisfy the boundary 
conditions in the different 
directions.  Screening functions or other approximations are thus not needed. 
For the direction with the free boundary conditions, the guidelines described in ref.\cite{freePS} are followed, 
properly adapted to the surfaces case.

The charge density in the non-periodic direction is represented using interpolating scaling functions, thus avoiding from the beginning long range interactions between super-cells.  
Like the free BC case, this Poisson solver is most efficient when dealing with localized densities (in the non-periodic direction). Such densities are for instance obtained from 
electronic structure codes using finite differences~\cite{chelikowsky} or finite elements~\cite{pask} or also Gaussians~\cite{quickstep} for the representation 
of the wave functions.

\section{Interpolating scaling functions}
Interpolating scaling functions \cite{lazy} arise in the framework of wavelet theory ~\cite{daub,sgbook}.
They are one-dimensional functions, and their three main properties are:
\begin{itemize}
\item The full basis set can be obtained from all the translations 
by a certain grid spacing $h$ of the mother function $\phi$ centered at the origin.
\item They satisfy the refinement relation
\begin{equation}
\label{refinement}
\phi(x) = \sum_{j=-m}^{m} \text{\sl h}_j \: \phi(2 x -j)
\end{equation}
where the $\text{\sl h}_j$'s are the elements of a filter that characterizes the wavelet family, and $m$ is the order of the scaling function.
Eq. \eqref{refinement} establishes a relation between the scaling functions on a grid with grid spacing $h$ and another one with spacing $h/2$. 
\item The mother function $\phi$ is symmetric, with compact support from $-m$ to $m$. It is equal to one at the origin and to zero at all other integer points (in grid spacing units). The expansion coefficients of any function in this basis are just the values of the function on the grid.
\end{itemize}

\section{Poisson equation for surfaces BC}
Consider a three-dimensional domain, periodic (with period $L_x$ and $L_y$) in $x$ and $y$ directions, and non-periodic in $z$. Without loss of generality, a function $f$ that lives in such a domain can be expanded as
\be\label{surface expansion}
f(x,y,z)=\sum_{p_x,p_y} e^{-2\pi i (\frac{p_x}{L_x} x + \frac{p_y}{L_y} y)} f_{p_x,p_y}(z)\;.
\ee
We indicate with $f_{p_x,p_y}(z)$ the one-dimensional function associated to the vector $\vec p=(p_x/L_x,p_y/L_y)$ in the reciprocal space of the two dimensional surface.
Following these conventions, the Poisson equation \eqref{poissonbase} become a relation between the reciprocal space components of $V$ and $\rho$:
\be\label{potentialmixed}
V_{p_x,p_y}(z)=-4 \pi \int_{-\infty}^{+\infty}\dd z'\, G(2\pi\,|\vec p|;z-z') \rho_{p_x,p_y}(z)\;,
\ee
where $|\vec p|^2=(p_x/L_x)^2+(p_y/L_y)^2$, and $G(\mu;z)$ is the Green function of the one-dimensional Helmholtz equation:
\be\label{greenhelmholts}
\left(\partial_z^2 - \mu^2\right) G(\mu;z) = \delta(z)\;.
\ee
The free BC on the $z$ direction fix the form of the Green function:
\be\label{formofgreen}
G(\mu; z)=
\begin{cases}
-\frac{1}{2 \mu} e^{-\mu |z|} & \mu > 0 \;,\\
\frac{1}{2} |z| & \mu = 0\;.
\end{cases}
\ee

In numerical calculations continuous charge distributions are typically represented by their values on a grid. The mixed representation of the charge density given above immediately suggests to use a plane wave expansion in the periodic directions, which may be easily treated with conventional FFT techniques.
For the non-periodic direction $z$ we will use interpolating scaling functions representation.
The corresponding continuous charge distribution is thus given by
\begin{multline}\label{mixedrep}
\rho(x,y,z)=\sum_{p_x=-\frac{N_x}{2}}^{\frac{N_x}{2}}\; \sum_{p_y=-\frac{N_y}{2}}^{\frac{N_y}{2}}\; \sum_{j_z=0}^{N_z} \rho_{p_x,p_y;j_z} \times \\ \times
\exp\left(-2\pi i (\frac{p_x}{L_x} x + \frac{p_y}{L_y} y)\right) \phi\left(\frac{z}{h}-j_z\right)\;,
\end{multline}
where $h$ is the grid spacing in the $z$ direction, and $\phi(j)=\delta_{j,0}$, $j \in \mathbb{Z}$.
This mixed representation of the charge density in eq.\eqref{mixedrep} has an important 
property. It is in a certain sense the most faithful continuous charge distribution 
that can be obtained from values on a grid. The main features of the electrostatic potential 
potential are determined by the multipoles and Fourier components of the charge distribution. 
For surface problems the multipoles are the significant quantity along the non-periodic 
direction and the Fourier components are the relevant quantities in the two periodic directions.
Using the completeness relation of plane waves (which holds also on a discrete real space grid) and the relation
\be
\int \phi(x-j) x^\ell = j^\ell\;,
\ee
which is demonstrated in \cite{freePS}, it is easy to show that 
\begin{multline}\label{multipoles}
 \int \dd x\, \dd y\, \dd z \cos\left(\frac{2\pi \ell_1 x}{L_x}\right)
\cos\left(\frac{2\pi \ell_2 y}{L_y}\right)
z^{\ell_3} \rho(x,y,z) = \\ =
\frac{L_x L_y h^{\ell_3+1}}{4}\!\!\!\sum_{\epsilon_{1,2}=\pm 1}\sum_{j_3=0}^{N_z}
j_3^{\ell_3} \rho_{\epsilon_1 \ell_1,\epsilon_2 \ell_2;j_3} = \\ =
L_x L_y h^{\ell_3+1} \sum_{j_1=0}^{N_x}\sum_{j_2=0}^{N_y}\sum_{j_3=0}^{N_z} 
\times\\ \times \cos\left(\frac{2\pi \ell_1 j_1}{N_x}\right)
\cos\left(\frac{2\pi \ell_2 j_2}{N_y}\right)
 j_3^{\ell_3}
 \rho(\frac{L_x}{N_x}j_1,\frac{L_y}{N_y}j_2,h j_3)\;,
\end{multline}
with $\ell_1 < N_x$, $\ell_2 < N_y$ and $\ell_3 < m$, where $m$ is the order of the scaling function.
This equation shows that the discrete multipoles and Fourier components of a charge density 
given on a grid are identical to the multipoles and Fourier components of the continuous 
$\rho$ given by eq.~\eqref{mixedrep}. For this reason it is not necessary that the input charge 
density for our method is given in the representation of eq.~\eqref{mixedrep}. It can, and it will 
actually in most cases, be given by numerical values on a grid. 

Combining eq. \eqref{potentialmixed} with \eqref{mixedrep}, the discretized Poisson problem thus becomes
\be \label{zconvol}
V_{p_x,p_y;j_z}=-4 \pi h \sum_{j'_z} K(2\pi\,|\vec p|;j_z-j'_z) \rho_{p_x,p_y;j'_z} \;,
\ee
where the quantity (kernel)
\be\label{kerneldefinition}
K(\mu;j) = \int_{-\infty}^{+\infty}\!\!\!\dd u\, G(\mu; h (j-u)) \phi(u)
\ee
is defined via an integral in the dimensionless variable $u$. Due to the symmetry of $\phi$, the kernel is symmetric in the non-periodic direction $K(\mu;j_z)=K(\mu;-j_z)$. The integral bounds can be restricted from $-m$ to $m$, thanks to the compact support of $\phi$.

Once we have calculated the kernel, which will be described below, our numerical procedure is the 
following. We perform a two-dimensional FFT on our real space charge density to obtain 
the Fourier coefficients $\rho_{p_x,p_y;j'_z}$ for all the periodic planes. 
Then we have to solve eq.~\eqref{zconvol}.
Since this equation is a convolution it can be calculated by zero-padded FFT's. 
Finally the potential is transformed back from the mixed representation to real space to obtain 
the potential on the grid by another two-dimensional FFT. 
Due to the FFT's, the total computational cost is $\mathcal O (N \log N)$. 
Since all quantities are real, the amount of memory  and the number of operations for the FFT 
can reduced by using real-to-complex FFT's instead of complex-complex FFT's. 

It remains now to calculate the values of the kernel function $K(\mu;j)$. 
The core of the calculation is represented by the function
\be\label{kappaintegral}
\tilde K(\lambda;j)=
\begin{cases}
\int \dd u\, e^{-\lambda |u-j|} \phi(u) &\lambda > 0\;,\\
\int \dd u\, |u-j| \phi(u) &\lambda = 0\;.
\end{cases}
\ee
The kernel has the properties $K(\mu;j)=-\tilde K(\mu h;j)/(2\mu)$ for $\mu>0$ and $K(0;j)= \tilde K(0;j)/2$.
A simple numerical integration with the trapezoidal rule is inefficient since $G(\mu;z)$ is not smooth in $z=0$ while the scaling function varies significantly around the integer points.
Thanks to the compact support of the scaling function, this problem can be circumvented with a simple and efficient recursive algorithm.
We define two functions $\tilde K^{(+)}$ and $\tilde K^{(-)}$ such that $\tilde  K(\lambda;j)=\tilde K^{(+)}(\lambda;j)+\tilde K^{(-)}(\lambda;j)$, where we have, for $\lambda>0$
\begin{align}
\tilde K^{(+)}(\lambda;j)& =\int_{-\infty}^{j} \dd u\, e^{\lambda (u-j)} \phi(u)\;, \label{beginrecursion}\\
\tilde K^{(-)}(\lambda;j)& =\int_{j}^{+\infty} \dd u\, e^{-\lambda (u-j)} \phi(u)\;,
\end{align}
while with $\lambda=0$
\begin{align}
\tilde K^{(\pm)}(0;j) &= \pm j\, Z_0^{(\pm)}(j) \mp Z_1^{(\pm)}(j)\;,\\
Z_\ell^{(+)}(j) &= \int_{-\infty}^{j} \dd u\, u^\ell \phi(u)\;,\\
Z_\ell^{(-)}(j) &= \int_{j}^{+\infty} \dd x\, u^\ell \phi(u)\;,\quad \ell=0,1\;.
\end{align}
These objects satisfy recursion relations:
\begin{align}
\tilde K^{(\pm)}(\lambda;j+1) &= e^{\mp \lambda} \left[\tilde K^{(\pm)}(\lambda;j) \pm e^{\mp \lambda j} D^{(\pm)}_\lambda(j)\right]\;,\notag\\
Z_\ell^{(\pm)}(j+1) &= Z_\ell^{(\pm)}(j) \pm C_\ell(j)\;,\quad \ell=0,1\;,
\end{align}
where 
\begin{align}
D^{(\pm)}_\lambda (j) &= \int_j^{j+1} \dd u\, e^{\pm \lambda u} \phi(u) \;,\\
C_\ell(j) &= \int_j^{j+1} \dd u\, u^\ell \phi(u) \;,\quad \ell=0,1\;.
\label{endrecursion}
\end{align}
From eq.(\ref{beginrecursion} -- \ref{endrecursion}), and the properties
\begin{align}
 \tilde K(\lambda;j)&=\tilde K(\lambda;-j)\;, &
 \tilde K^{(+)}(\lambda;0)&=\tilde K^{(-)}(\lambda;0)\;, \\
 Z_1^{(+)}(0)&=Z_1^{(-)}(0)\;, &
 Z_0^{(+)}(0)&=Z_0^{(-)}(0)=\frac{1}{2}\;,\notag
\end{align}
the function $\tilde K(\lambda;j)$ can be calculated recursively for each $j \in \mathbb{N}$, by knowing
 $\tilde K^{(+)}(\lambda;0)$ and $Z_1^{(+)}(0)$, then evaluating $D^{(\pm)}_{\lambda} (j)$ and $C_\ell(j)$ for each value of $j$.
The integrals involved can be calculated with high accuracy with a simple higher-order polynomial quadrature. They are integral of smooth, well-defined quantities, since the interpolating scaling function goes to zero at their bounds. Moreover, for values of $j$ lying outside the support of $\phi$ we can benefit of a functional relation for calculating the values of the kernel.
The support of a $m$-th order scaling function goes from $-m$ to $m$, then we have $\forall p >0$
\begin{align}
 K(\mu;m+p)&=e^{-\mu h p} K(\mu;m)\;,\quad \mu>0\;,\notag\\
 K(0;m+p)&=K(0;m)+p\, Z_0^{(+)}(m)\;. \label{overborder}
\end{align}
To summarize, we have found an efficient method for evaluating equation \eqref{kerneldefinition} for $j=0,\cdots,N_z$ and a fixed $\mu$.
Instead of calculating $N_z+1$ integrals of range $2m$, we can obtain the same result by calculating 2 integrals of range $m$ and $4m$ integrals of range 1, with the help of relation \eqref{overborder}. This will also increase accuracy, since the integrands are always smooth functions, which would not be the case with a naive approach.

The accuracy in calculating the integrals can be further improved by
using the refinement relation \eqref{refinement} for interpolating scaling functions.
For positive $\lambda$ we have
\begin{align}
 \tilde K(2 \lambda;i)& = \int \dd u\, e^{- 2 \lambda |u-i|} \phi(u) \notag \\
& = \frac{1}{2} \int \dd u\, e^{- \lambda |u-2 i|} \phi(u/2) \notag \\
& = \frac{1}{2} \sum_j \text{\sl h}_j \int \dd u\, e^{- \lambda |u-2 i|} \phi(u - j)  
\label{recursion}\\
& = \frac{1}{2} \sum_j \text{\sl h}_j \tilde K(\lambda;2 i - j)\;.\notag
\end{align}
This relation is useful to improve the accuracy in evaluating the kernel for high $\lambda$.
Since in this case the exponential function is very sharp, it is better to calculate the kernel for lower $\lambda$ and an enlarged domain and then apply relation \eqref{recursion} as many times as needed.
The relation \eqref{overborder} allows us to enlarge the domain with no additional computational cost.
With the help of the above described properties the computational time for evaluating the kernel in Fourier space can be considerably optimized, becoming roughly half of the time needed for its application on a real space density.

\section{Numerical results and comparison with other methods}
Our new method has an algebraic convergence rate of $h^m$, where $h$ is the grid spacing and 
$m$ is the order of the interpolating scaling function used. This is due to the fact that 
the charge density along the non-periodic direction can be represented with an approximation 
error of the order of $h^m$ using interpolating scaling functions of order $m$~\cite{sgbook}. 
The error of the plane wave representation of the charge density in the periodic plane has 
a faster exponential convergence rate and is thus not visible in the overall convergence rate. 
By choosing higher order interpolating scaling function we can get arbitrarily high 
algebraic convergence rates.
Our method was compared with the reciprocal space methods by Hockney~\cite{Hockney Surfaces}
and Mortensen~\cite{Mortensen Surfaces} as implemented in the CPMD electronic structure program~\cite{cpmd}.
\begin{figure}[htbp]
\includegraphics[width=.45\textwidth]{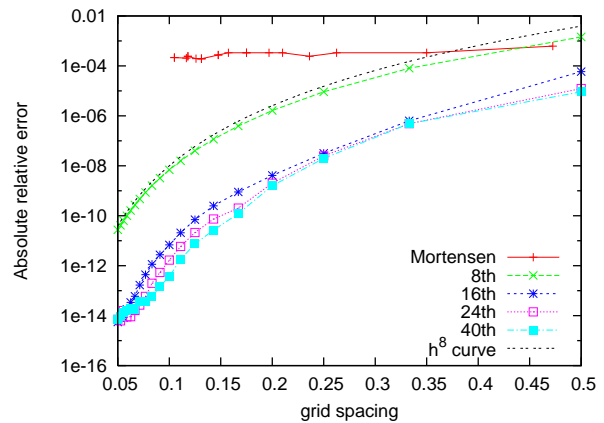}
\caption{Accuracy comparison between our method with scaling functions of different orders 
and the Mortensen solver for surface systems as implemented in CPMD. 
The results of the Hockney method are not shown since they are much less precise. 
The $h^8$ curve is plotted to show the algebraic decrease of the precision with respect to  the grid space $h$. The accuracy is finally limited by the evaluation of the integral \eqref{kappaintegral}, which is computed with nearly machine precision.}
\label{accuracy}
\end{figure}
The tests shown in figure \ref{accuracy} are performed with an analytical charge distribution that is the Laplacian of $V(x,y,z)=\exp(\cos(\frac{2\pi}{L_x} x)+\cos(\frac{2\pi}{L_y} y))\exp(-\frac{z^2}{50 L_z^2} - \tan(\frac{\pi}{L_z} z)^2)$. 
Its behavior along the $xy$ surface is fully periodic, with all the reciprocal space components taken into account. The $\exp(- \tan(\frac{\pi}{L_z} z)^2)$ guarantees a localized behavior in the non-periodic direction with the potential going explicitly to zero at the borders. This makes also this function suitable for comparison with reciprocal space based approach. The Gaussian factor is added to suppress high frequency components. 
Tests with other analytical functions gave comparable accuracies.
The reciprocal space Poisson solvers turn out to be much less precise than our approach, which explicitly preserves the BC along each direction. Moreover, the accuracy shown for the Mortensen approach is calculated only for planes that lies in the bulk of the non-periodic direction (30\% of the total volume). Outside of this region, errors blow up, which will imply that a very large computational volume is needed to obtain accurate results in a sufficiently large domain of the non-periodic direction.

\begin{figure}[htbp]
\includegraphics[width=.45\textwidth]{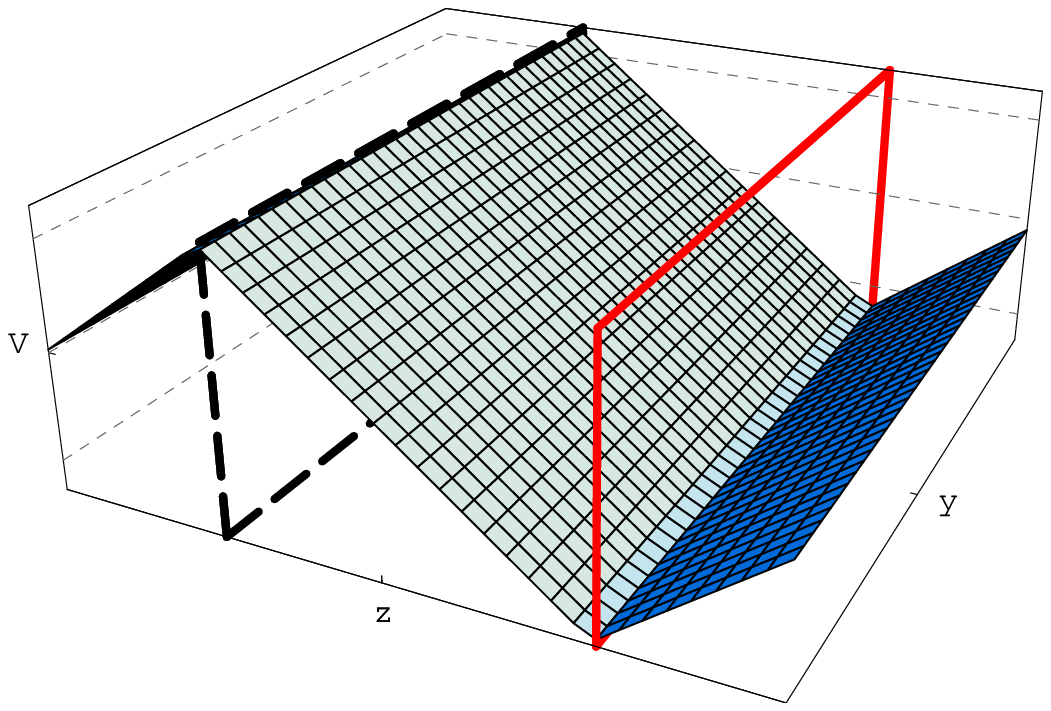}
\includegraphics[width=.45\textwidth]{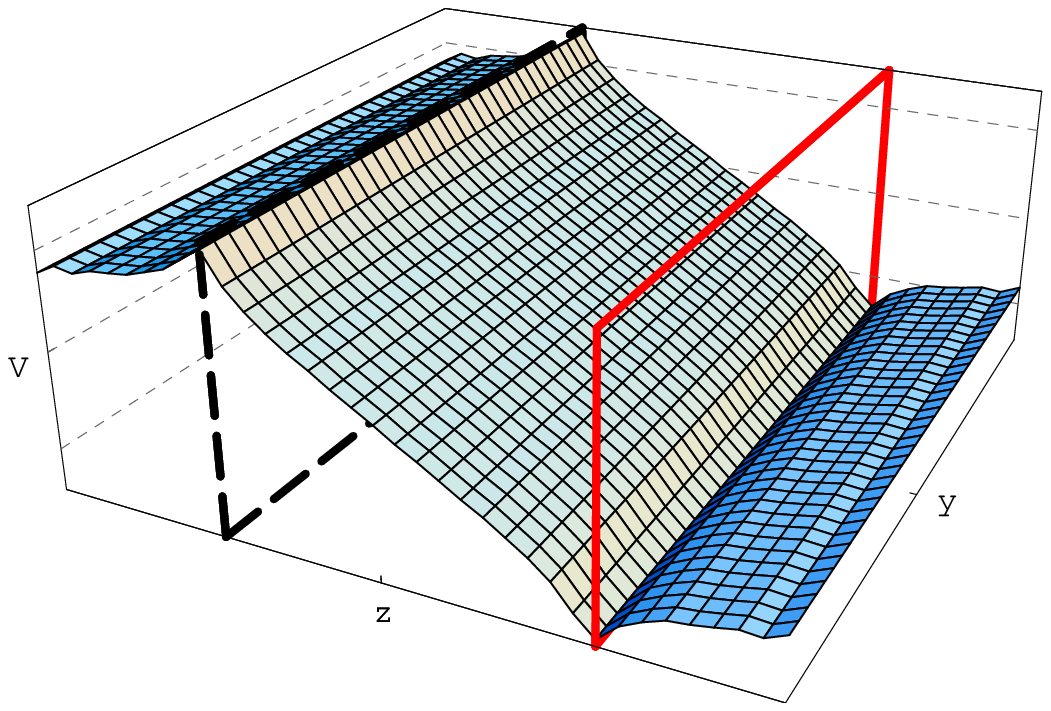}	
\includegraphics[width=.45\textwidth]{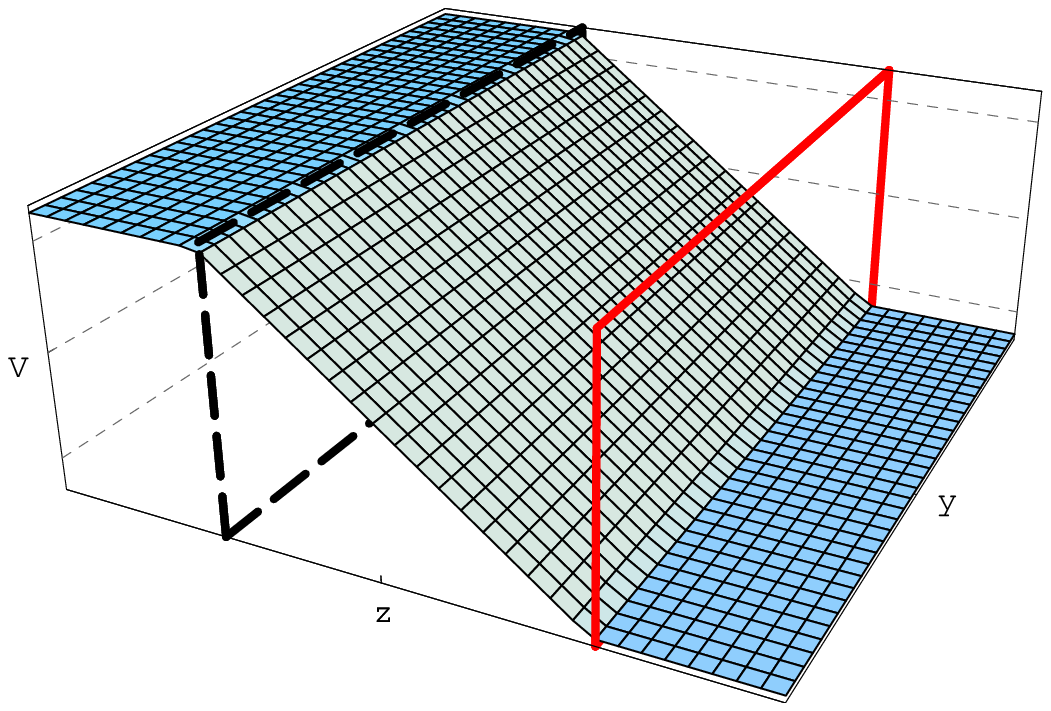}
\caption{Electrostatic potential $V$ for a system with two periodic planes charged with opposite sign (plane capacitor), oriented along the $z$ direction, calculated by different Poisson solvers. The values of $V$ are taken in the middle of the $x$ (periodic) direction. The position of the positive (black, dashed line) and the negative (red, solid) charged plane is schematically shown in the figure. The represented solution are, from top to bottom, the results from the Mortensen, the Hockney and our Poisson solver.}
\label{surfaces}
\end{figure}

To show that our method genuinely preserves the boundary conditions appropriate for surfaces
we calculated the electrostatic potential for a plane capacitor. For this system only the 
zero-th Fourier components in the plane are non-vanishing. 
Figure \ref{surfaces} shows the results either in the Mortensen/ Hockney reciprocal space 
methods or with our approach.
For the plane capacitor, the screening function used in the Mortensen approach vanishes, and the solution 
is equal to what we would have obtained with a fully periodic boundary conditions. 
To obtain the good ``zero electric field'' behavior in the borders that we obtain directly with 
our method one would have to postprocess the solution obtained from the Mortensen method, 
by adding to the potential a suitable linear function along the non-periodic direction. 
This is legitimate since a linear 
function is annihilated by the Laplace operator and the modified potential is thus 
also a valid solution of the Poisson equation just with different boundary conditions. 
The Hockney method presents a better qualitative behavior, though the results are not accurate.
Only with our approach we get both accurate and physically sound results.

Table~\ref{timings} shows the required CPU time for solving the Poisson equation on a grid of 
$128^3$ grid points as a function of the number of processors on a Cray XT3 parallel computer. 
The parallel version is based on a parallel 3-dimensional FFT, where the input/output is properly distributed/gathered to all the processors. The FFT's are performed using a modified version of the algorithm described in ref.\cite{steFFT} that gives high performances on a wide range of computers.
\begin{table}[htbp]
\begin{tabular}{cccccccc} 
\hline
\hline
  1 & 2 & 4 & 8 & 16 & 32 & 64 & 128  \\ 
\hline 
.43 & .26 & .16 & .10 & .07 & .05 & .04 & .03 \\
 \hline 
%.37 & .32 & .22 & .19 & .15 & .14 & .16 & .16 \\ itanium CEA
%.37 & .30 & .18 & .10 & .06 & .04 & .05 & .05 \\ \hline  itanium CEA distributed
\hline
\end{tabular}
\caption{ The elapsed time in seconds required on a Cray XT3 (based on AMD Opteron processors) to solve Poisson's equation with surface BC
on a 128$^3$ grid as a function of the number of processors. The time for setting up the Kernel (around 50\% of the total time) is not included.
For a large number of processors, the communication time needed to gather the complete potential to all the processors becomes dominant.}
\label{timings}
\end{table}

 A package for solving Poisson's equation in real space according to the method described here 
 can be downloaded from http://www.unibas.ch/comphys/comphys/SOFTWARE

In conclusion, we have presented a method that allows us to obtain accurate potentials arising from charge 
distributions on surfaces with a $O(N \log N)$ scaling in a mathematically clean way. This method preserves explicitly the required boundary conditions, and can easily be used for applications inside electronic structure codes where the charge density is either given in reciprocal or 
in real space. 

We acknowledge interesting discussions with Fran\c{c}ois Bottin and Jaroslaw Piwonski.
 This work was supported by the European Commission within the Sixth Framework Program through NEST-BigDFT (contract no. BigDFT-511815) and by the Swiss National Science Foundation.
 The timings were performed at the CSCS (Swiss Supercomputing Center) in Manno. 
% 


\end{document}